\documentclass[conference, letterpaper]{IEEEtran}
\usepackage{cite}
\usepackage{amsmath,amssymb,amsfonts,amsthm}
\usepackage{algorithmic}
\usepackage{textcomp}
\usepackage{xcolor}
\usepackage{graphicx}
\usepackage{listings}
\usepackage{mdframed}
\usepackage[font=small]{caption}
\usepackage{subcaption}
\usepackage{array}
\usepackage{booktabs}
\usepackage{multirow}
\usepackage{makecell}
\usepackage{dblfloatfix}
\usepackage{blindtext}
\usepackage{hyperref}

\newcommand{\rulesep}{\unskip\ \vrule\ }

\newtheorem{definition}{Definition}

\def\BibTeX{{\rm B\kern-.05em{\sc i\kern-.025em b}\kern-.08em
    T\kern-.1667em\lower.7ex\hbox{E}\kern-.125emX}}
\begin{document}

\title{Specognitor: Identifying Spectre Vulnerabilities via Prediction-Aware Symbolic Execution}

\author{\IEEEauthorblockN{Ali Sahraee}
\IEEEauthorblockA{ \textit{Linköping University}\\
Linköping, Sweden \\
ali.sahraee@liu.se}
}

\maketitle

\begin{abstract}
Spectre attacks exploit speculative execution to leak sensitive information. In the last few years, a number of static side-channel detectors have been proposed to detect cache leakage in the presence of speculative execution. However, these techniques either ignore branch prediction mechanism,  detect static pre-defined patterns which is not suitable for detecting new patterns, or lead to false negatives.

In this paper, we illustrate the weakness of prediction-agnostic state-of-the-art approaches. We propose Specognitor, a novel prediction-aware symbolic execution engine to soundly explore program paths and detect subtle spectre variant 1 and variant 2 vulnerabilities. We propose a dynamic pattern detection mechanism to account for both existing and future vulnerabilities. Our experimental results show the effectiveness and efficiency of Specognitor in analyzing real-world cryptographic programs w.r.t. different processor families.
\end{abstract}


\section{Introduction}
\label{sec:intro}

The performance of modern processors have increased significantly during past decades by increasing the working clock frequency and shrinking transistor technology. However, they reached their performance limit due to increased fault rate and increased temperature. To further increase the CPU performance, vendors applied other performance enhancement techniques such as increasing the number of cores and optimizing the instruction pipelining.

Instruction pipelining includes complex parallelization techniques to increase the processor performance. However, unresolved dependency issues of instructions could stall the pipelines and reduce the instruction-level parallelism. To overcome the dependency issues, modern processors use speculative execution \cite{tomasulo1967efficient}. During speculative execution, the processor uses micro-architectural optimizations to predict the control flow dependencies, data dependencies and even the actual data. CPU prefetch units employ techniques to predict future instruction stream based on previous patterns. The execution continues according to the prediction until the prediction is evaluated. Since the prediction may turn out to be wrong (e.g., mis-predicted branch outcome), the processor might need to \textit{roll back} the execution in which it discards the executed instructions, retrieves the last valid micro-architectural state (i.e., registers and buffers including PC), and re-fills the CPU piplelines with the correct instruction stream. 

Although the correct architectural state of the system is restored while rolling back, the micro-architectural side effects of the speculative execution remains in the system, e.g., loaded cache lines. Based on these side effects, the family of Spectre \cite{kocher2020spectre} attacks have been discovered. Spectre attacks use the side-effects of speculative execution to obtain sensitive information. Unfortunately, detecting Spectre attacks is challenging since there are many factors that can affect the system behavior and the length of speculative execution such as the processor speed, the concurrency level, the memory link speed, and etc. The most precise approach for detecting Spectre-type vulnerabilities in a program is to run it on the physical system and to try one input at a time and monitor the micro-architectural side effects of speculation during execution. This approach is extremely slow because of limited hardware resources, large number of inputs, and large number of factors that can affect the execution \cite{muench2018you}. Other approaches apply offline techniques to run the code with respect to (w.r.t.) abstractions about the environment and using symbolic values as input. Symbolic execution  \cite{king1976symbolic, avgerinos2014enhancing} based techniques have been traditionally used to explore program paths (w.r.t. path conditions) to find software vulnerabilities \cite{bounimova2013billions, cha2012unleashing}, to generate high coverage test cases \cite{cadar2008klee}, and to find inputs that cause program errors \cite{cadar2008exe, godefroid2005dart}. Therefore, they mainly do not consider the micro-architectural components (e.g., prediction logic) in their analysis. However, recent approaches \cite{brotzman2021specsafe, guarnieri2020spectector, guo2020specusym, wang2020kleespectre} employ a non-deterministic and history-agnostic prediction logic only for conditional branch instructions. These approaches apply inaccurate and inefficient exploration of the program execution states that results in detecting false positive and false negative cases. These approaches are inaccurate because they do not consider Branch Target Buffer (BTB) for conditional branch prediction while some processors (e.g., Intel Pentium 4 \cite{hinton2001microarchitecture}) only employ BTB in their prediction logic as the primary means of prediction. These approaches are also inefficient because they do not consider an accurate prediction logic for speculative execution which results in many infeasible path exploration. For example, our evaluation results show up to 80\% additional infeasible speculative path exploration by \textsc{KLEESpectre} \cite{wang2020kleespectre}.


In addition, some techniques \cite{brotzman2021specsafe, guarnieri2020spectector, guarnieri2021hardware} that employ non-interference in their analysis lack the ability to record the system state along the program execution. For example, SpecSafe \cite{brotzman2021specsafe} asks a solver to generate any history trace for prediction logic that could cause information leakage. This approach considers a powerful adversary that can interrupt the victim process and train the branch prediction at any moment, while modern processors \cite{arm2018cortexa53, arm2012cortexa9, arm2013cortexa7} mainly use process-specific information for branch prediction which can be flushed on a context switch. In other words, training the branch prediction unit by adversary might not affect the branch prediction outcome of the victim process. Hence, considering all prediction outcomes results in capturing false positives for many processor families. In addition, recent approaches that employ speculation do not address indirect branch prediction.


We propose Specognitor an accurate and efficient prediction-aware symbolic execution framework to soundly explore all feasible program paths and search for information leakage w.r.t. a certain processor. Specognitor simulates the micro-architectural state of the system during symbolic execution, generates the possible speculative paths according to the prediction logic, and searches for certain patterns and information leakage. Our framework explores fewer speculative paths than existing techniques \cite{wang2020kleespectre, brotzman2021specsafe, guarnieri2020spectector} because of the underlying prediction logic. However, since it employs a dynamic prediction model, it might explore completely different speculative paths than existing approaches (e.g., in case of using Branch Target Buffer). 


To sum up, this paper makes the following contributions:\\
\noindent $\bullet$ We show inaccuracy of the state-of-the-art tools in detecting spectre vulnerabilities w.r.t. the micro-architectural configuration of a system.\\
\noindent $\bullet$ We propose Specognitor\footnote{Specognitor is available at \url{https://github.com/sahraeeali/Specognitor}}, a prediction-aware symbolic execution framework to automatically explore program paths and detect speculative cache side channels. \\
\noindent $\bullet$ We propose a pattern detection monitor to effectively define security threats and find them in execution traces.\\
\noindent $\bullet$ We incorporate taint analysis in symbolic execution to propagate meta information.\\
\noindent $\bullet$ We evaluate efficiency, real-world applicability, and scalability of Specognitor to detect speculative cache side-channels w.r.t. spectre variant 1 and variant 2.

\textbf{Outline.} This paper is organized as follows. Section~\ref{sec:background} overviews the technical background of Specognitor. Section~\ref{sec:predictionaware} outlines the details of prediction-aware symbolic execution and the pattern detection monitor. Section~\ref{sec:implementation} provides technical implementation details of Specognitor. In section~\ref{sec:evaluation}, we present our evaluation results. In section~\ref{sec:discussion}, we further discuss features, limitations, and future works. In section~\ref{sec:relatedwork}, we review the related work. We conclude our work in Section~\ref{sec:conclusion}.



\section{Backgound}
\label{sec:background}








Before getting into more details of Specognitor, we discuss the technical backgrounds that we use for implementing it.

\subsection{Prediction Logic}
Modern processors use branch predictors to increase the number of instructions available for execution that helps in utilizing instruction level parallelism (ILP). A branch direction predictor speculates whether a conditional branch is likely to be taken or not. A branch target predictor predicts the target of branch/call instructions. These units can employ static or dynamic techniques. Static branch prediction techniques are simple and easy to implement and require little hardware resources. These techniques do not use any run-time information about branch behavior. However, dynamic branch prediction techniques are more complex and use the run-time information to enhance their performance. These techniques can adapt to the behavior of the program during execution \cite{kulkarni2016review}. In this paper, we use two common prediction techniques for direction prediction and target address prediction. As we see in section ~\ref{subsec:realExample}, most processors use these two techniques to implement their prediction logic.
\subsubsection{Two-Level Prediction Table}
\label{subsec:PHT}
The two-level prediction table is a branch direction prediction technique that uses two separate levels of information to predict the branch outcome.

The first level consist of the history of the last k branches stored in a \textit{Branch History Register} (BHR). BHR is an n-bit shift register that records "1" if a branch is taken and "0" if it is not taken. The BHR shifts out the information of oldest branch when the outcome of a new branch is available. The second level of this prediction technique is called \textit{Pattern History Table} (PHT). PHT is a $2^m$-entry table of saturating counters. There are three implementation schemes for each level, i.e., global history scheme, per-set history scheme, and per-address history scheme. The implemented scheme determines the number of BHRs and PHTs, the number of address bits used (from PC), and the relation between $n$ and $m$. Elaborate details of these schemes are contained in \cite{yeh1992alternative, yeh1993comparison, pan1992improving}.
In this paper, we use the global history scheme for both levels parameterized by $n$. In other words, we use an $n$-bit BHR and a $2^n$-entry PHT of 2-bit saturating counter. When the actual result of the branch is evaluated, its respective counter is incremented in case of taken or decremented in case of not taken. Since we use the global history scheme, we do not use the PC value. Figure~\ref{img:PHT} shows a two level prediction table in which both levels use global history scheme. The content of BHR is $0111$ which is used to index the PHT. The 2-bit saturating counter of the PHT entry shows the value of $10$ which means the prediction outcome is \textit{taken}.


\subsubsection{Branch Target Buffer}
\label{subsec:BTB}
\begin{figure}
\centering
\begin{subfigure}{.2\textwidth}
  \centering
  \includegraphics[width=\textwidth, height=0.23\textheight]{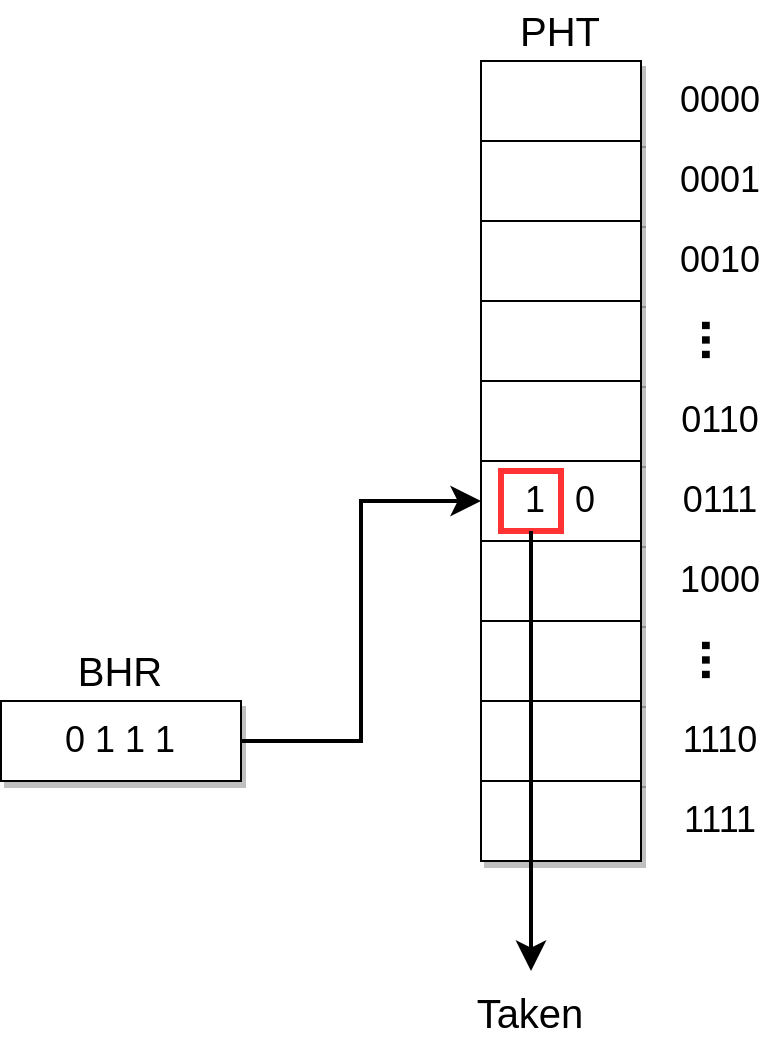}
  \caption{\footnotesize Two-level predictor with a 4-bit global history BHR and a 16-entry global history PHT.}
  \label{img:PHT}
\end{subfigure}%
\rulesep
\begin{subfigure}{.25\textwidth}
  \centering
  \includegraphics[width=\textwidth]{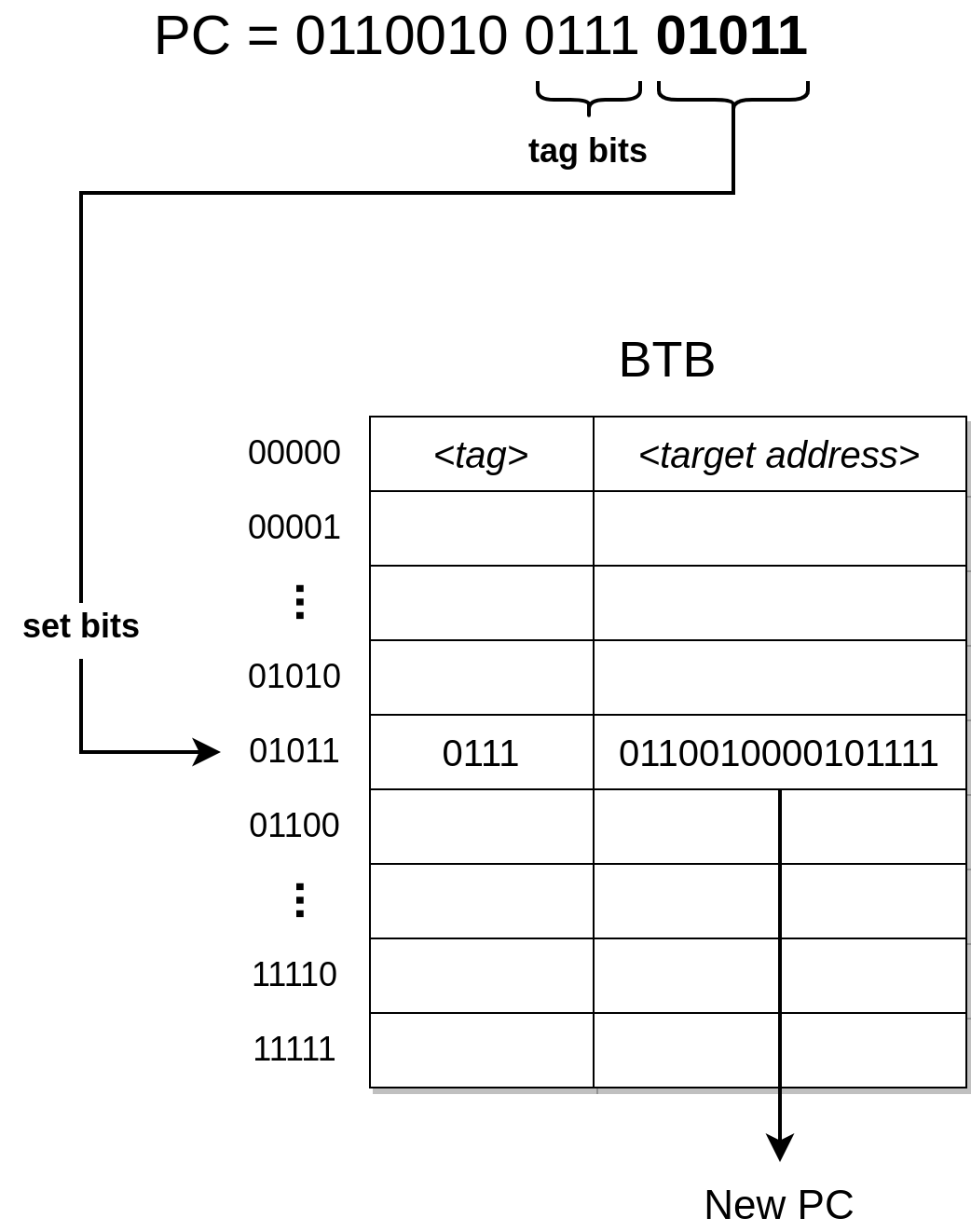}
  \caption{\footnotesize 32-entry direct-mapped BTB with four bits of tag implemented in per-set history scheme.}
  \label{img:BTB}
\end{subfigure}
\caption{\small Two examples of branch prediction units.}
\label{img:PredictionLogic}
\end{figure}

The \textit{Branch Target Buffer} (BTB) is a component that stores the target address of branch instructions. BTB architecture is close to cache with sets, tag bits, and different set-associativities. BTB can also be implemented in global history scheme, per-set history scheme, and per-address history scheme. These schemes determine the number of sets and the number of bits used as tag. Figure~\ref{img:BTB} shows the architecture of a 32-entry direct-mapped BTB (BTB:) with 4 bits of tag implemented with per-set scheme \cite{yeh1993comparison}. While speculating, if the branch predictor predicts the branch to be taken and there is a hit in the BTB (BTB-hit), the stored address in the respective BTB set is used as the next PC. If the branch predictor predicts the branch to be taken and the respective entry is not in the BTB (BTB-miss), the processor can stall the execution or behave as if the branch is not taken. When the prediction outcome is resolved, the respective BTB-entry is updated.

\begin{figure}
\centering
\includegraphics[scale=0.14]{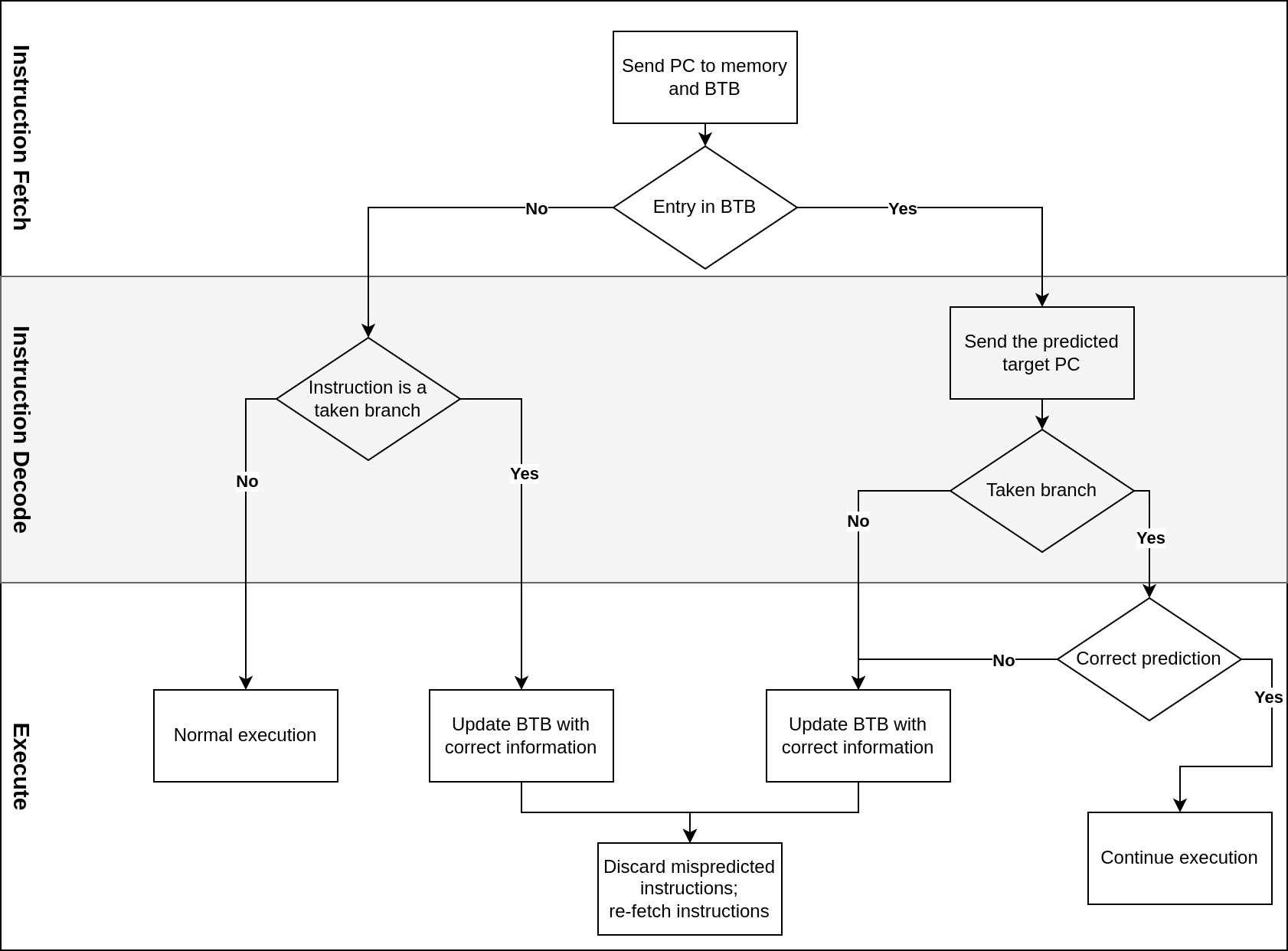}
\caption{\small Workflow of branch target buffer through different pipeline stages.}
\label{img:BTBworkflow}
\end{figure}


\subsection{Branch Predictors in modern Processors}
\label{subsec:realExample}

In this section, we study the branch prediction mechanisms implemented by different architectures.

\subsubsection{ARM - Cortex-A53}
\cite{arm2018cortexa53} is a low-power, mid-range processor that implements the ARMv8-A architecture instruction set. It is an in-order 8-stage pipeline processor with symmetric dual-issue unit. The Instruction Fetch Unit of Cortex-A53 processor contains a 2-way set associative cache that can hold 16 instructions (32 16-bit T32 instructions). It uses two prediction schemes. The target predictor contains a 256-entry BTB to predict the target address of indirect branches/calls. The direction predictor is implemented according to the global prediction scheme that uses BHRs and a 3072-entry PHT.

\subsubsection{ARM - Cortex-A7}
\cite{arm2013cortexa7} is a high-performance, low-power processor that implements ARMv8-A architecture instruction set. It is an in-order 8-stage pipeline processor with direct and indirect prediction units \cite{shimpi2022arm}. It has a 2-way set associative instruction cache that can hold 8 ARM instructions (16 Thumb 16-bit instructions). Its target prediction unit contains an 8-entry BTB (BTAC) and its direction prediction unit adopts a global prediction scheme that uses BHRs and a 256-entry PHT.

\subsubsection{ARM - Cortex-A9}
\cite{arm2012cortexa9} is a high-performance, low-power, out-of-order ARM processor that implements ARMv7 architecture instruction set. It is a multi-issue superscalar, 8-stage pipeline processor that predicts all branch instructions, such as conditional, unconditional, and indirect branches regardless of the addressing mode. The prediction logic contains both target prediction and direction prediction. The target prediction unit uses a 2-way BTB (BTAC) with 512, 1024, 2048, or 4096 entries. The direction prediction unit uses a global history scheme that has 1024, 2048, 4096, 8192 or 16384 PHT entries. 

\subsubsection{Alpha - 21264}
\cite{kessler1999alpha} is a high-speed, superscalar, 7-stage pipeline, out-of-order processor. It accepts 80 in-flight instructions. Its branch prediction unit uses a complex technique which implements two separate branch predictors and dynamically chooses one of their outcomes to fetch instructions. The first technique uses a two-level prediction with local history. Its first level comprises a table that holds 10 bits of history for up to 1024 different branches. The 10-bit history is then used to index a 1024-entry PHT. The second technique uses a global prediction scheme where a 12-bit BHR keeps the history of the last 12 branches and is used to index a 4096-entry table of 2-bit saturating counters.

\subsubsection{Intel - Pentium$^{\textregistered}$ 4}
\label{subsub:p4}
\cite{hinton2001microarchitecture} is a high-performace processor manufactured by Intel based on NetBurst microarchitecture \cite{koufaty2003hyperthreading}. It is an out-of-order, 20-stage pipeline processor with branch prediction unit. The dynamic branch predictor uses a 4096-entry BTB to capture the branch history information of the program. In case of BTB-miss, it uses a simple static \textit{Backwards Taken Forward Not Taken} (BTFNT) \cite{intel2003refop} branch prediction technique.

\subsubsection{ARM 11}
\cite{arm2008arm11} is a high-performance, low-power, 8-stage pipeline processor that implements ARMv6 architecture instruction set. It uses two branch prediction techniques. The dynamic branch predictor contains a 128-entry BTB (BTAC) to predict the address of a branch instruction. In case of BTB-miss, the prediction unit uses the BTFNT \cite{kulkarni2016review, intel2003refop} technique.

\subsection{Side-Channel Attacks}
Side-channel attacks \cite{kocher1996timing} exploit the side-effects of computation from the physical implementation of a system to infer sensitive information. These side-effects include execution time of cryptographic algorithms \cite{kocher1996timing}, power and energy consumption of the system \cite{kocher1999differential}, electromagnetic emissions\cite{agrawal2002side, longo2015soc}, and computation faults \cite{joy2011side}. However, cache attacks \cite{bernstein2005cache, schwarz2019netspectre, osvik2006cache} are the most prominent type of side-channel attacks.

Cache attacks measure the access time of different memory locations to reconstruct the secret information. The CPU cache is a small but fast memory inside the processor which reduces memory access time by storing the frequently accessed data. Different types of cache attacks are Evict+Time \cite{osvik2006cache, gruss2015cache, lipp2016armageddon}, Prime+Probe \cite{osvik2006cache, percival2005cache, bonneau2006cache, liu2015last, tromer2010efficient}, and Flush+Reload \cite{yarom2014flush, gruss2016flush}.

Flush+Reload attacks rely on shared cache-line granularity. In this type of side-channel attacks, the adversary flushes the last-level cache and tries to reload the shared data. Small delay in reloading the data shows that the victim has already loaded the cache line into CPU cache. On the other hand, Prime+Probe attacks does not rely on shared memory. The adversary fills the CPU cache and checks whether the victim evicts any cache entry from the cache.

\subsection{Spectre Attack}
\label{subsec:spectre}
Spectre \cite{kocher2020spectre} is a side-channel vulnerability has been discovered in modern processors. Spectre-style attacks deliberately exploit speculative execution in victim's code to leak the content of an arbitrary memory location, for example, the secret key of a cryptographic computation. Kocher et al. \cite{kocher2020spectre} categorized Spectre vulnerabilities into 4 variants w.r.t. the poisoned architectural component. These variants are as follows: i) in \textit{spectre variant 1}, the adversary uses conditional branches as the source of misprediction; ii) in \textit{spectre variant 2}, the adversary uses mispredicted indirect branches/calls to leak information; iii) in \textit{spectre variant 3}, the adversary poisons \textit{Return Address Buffer} (RSB) to mis-direct the execution flow; and iv) in \textit{spectre variant 4}, the adversary uses the memory disambiguation unit to mispredict data flow dependencies.



In this paper, we mainly focus on Spectre variant 1 and variant 2 as they target branch prediction unit of the processor. 

\begin{figure}
\centering
\begin{subfigure}{.3\textwidth}
  \centering
    \begin{lstlisting}[language=C,numbers=left,stepnumber=1,frame=lines,basicstyle=\ttfamily\scriptsize,xleftmargin=3em,framexleftmargin=2em]
uint8_t victim_fun(uint32_t x)
{
    uint8_t temp = 0;
    if (x < array1_size) {    
        temp = array1[x];
        temp &= array2[temp*64];
    }
    return temp;
}
  \end{lstlisting}
  \caption{\footnotesize Example code.}
  \label{img:01}
\end{subfigure}%
\begin{subfigure}{.2\textwidth}
  \centering
  \includegraphics[scale=0.1]{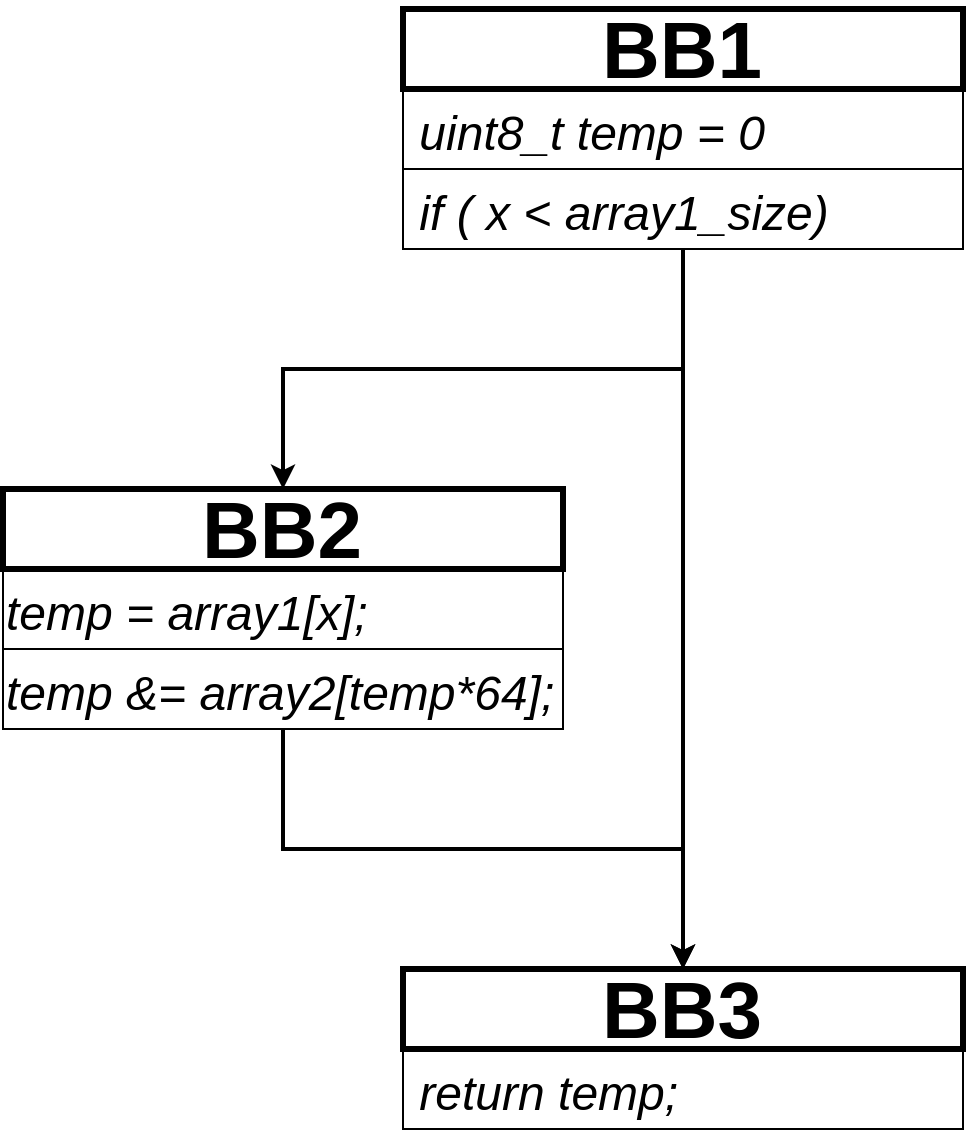}
  \caption{\footnotesize Control flow graph.}
  \label{img:cfg1}
\end{subfigure}
\caption{Traditional Spectre variant 1 example and its control flow graph.}
\label{img:Example1}
\end{figure}

\noindent \textbf{Spectre Variant 1.} Figure~\ref{img:01} represents a code fragment taken from \cite{kocher2020spectre} which demonstrates a conventional spectre variant 1 vulnerability. In this example, assume array1 and array2 both contain only public data. Moreover, assume the prediction unit predicts the condition in line 4 as taken while $x > array1\_{}size$. Assume the adversary can control the value of $x$ and tries to learn the secret key which is placed somewhere in the victim's memory space. According to our assumptions, the adversary can access an out-of-bound value (i.e., the key) by controlling the value of $x$ in $array1[x]$. Loading the secret key occurs during speculative execution and the instruction will be discarded after evaluating the condition at line 4, so it does not violate the correctness of the program. However, execution of out-of-bound memory access and its subsequent memory access at line 6 would result in loading a key-dependent cache line into the CPU cache. The adversary can infer the value of the secret key by performing a conventional cache attack. The vulnerability in Figure~\ref{img:01} is based on a specific pattern description: i) a vulnerable branch instruction (line 4), ii) a load secret instruction (line 5), and iii) a load instruction which expose the key to the adversary (line 6). In this paper, we refer to this pattern as {\ttfamily BR-LD-LD}. However, recent work \cite{guarnieri2021hardware, nemati2020speculative, brotzman2021specsafe} has shown other patterns categorized as spectre variant 1. Figure ~\ref{img:02} shows a code fragment where loading the secret value happens before the branch instruction, thus it could be represented as {\ttfamily LD-BR-LD} pattern.

\begin{figure}
\centering
    \begin{lstlisting}[language=C,numbers=left,stepnumber=1,frame=lines,basicstyle=\ttfamily\scriptsize,xleftmargin=3em,framexleftmargin=2em]
uint8_t victim_fun(uint32_t x)
{
    uint8_t temp = array1[x];
    if (x < array1_size) {    
        temp &= array2[temp*64];
    }
    return temp;
}
  \end{lstlisting}
  \caption{Another Spectre variant 1 example.}
  \label{img:02}
\end{figure}

\noindent \textbf{Spectre Variant 2.} Figure~\ref{img:03} shows a code example categorized as spectre variant 2 \cite{canella2019systematic}. In this example, assume that the BTB entry for the indirect call at line 10 contains the address of {\ttfamily\small unsafe\_{}function}. In other words, when the processor reaches line 10 and asks for a target from BTB, the BTB returns the address of {\ttfamily\small unsafe\_{}function}. However, according to line 16, the {\ttfamily\small safe\_{}function} should be executed. Further, assume the adversary can control the value of $idx$. While the processor speculatively executes the {\ttfamily\small unsafe\_{}function}, the instruction at line 6 loads an out-of-bound secret key and the instruction at line 7 exposes the secret to cache attacks.

\begin{figure}
\centering
    \begin{lstlisting}[language=C,numbers=left,stepnumber=1,frame=lines,basicstyle=\ttfamily\scriptsize,xleftmargin=3em,framexleftmargin=2em]
void safe_func(uint32_t x){
  // Safe computation
}
void unsafe_func(uint32_t x)
{
  temp = array1[x];
  temp &= array2[temp*64];
}
uint32_t ind_call(uint8_t i, uint32_t val){
  uint32_t res = (*p[i]) (val);
  return res;
}
void victim_fun(uint32_t idx){
  p[0] = safe_func;
  p[1] = unsafe_func;
  ind_call(0, idx);
}
  \end{lstlisting}
  \caption{Spectre variant 2 example.}
  \label{img:03}
\end{figure}

A few defense mechanisms have been proposed to protect the system against spectre side-channel attacks variant 1 and variant 2. Two of the most prevalent techniques used for spectre variant 1 and variant 2 is fencing \cite{kocher2018fence} and Retpoline \cite{kadir2019retpoline}, respectively. Placing fence instructions before branches disables the speculation which increases the execution time significantly. Moreover, applying Retpoline increases both execution time and energy consumption since the processor executes some dummy instructions. Compilers use these mechanisms to secure a program against spectre attacks. However, they blindly apply them for all branch instructions in the program. To reduce the cost of these mechanisms, static analysis tools try to optimize usage of defense mechanisms to reduce the execution time and energy consumption of the program while maintaining the system safety and security.

\subsection{Threat and Adversary Model}
Here, we consider that the victim and attacker codes are running on the same physical device. We assume our processor is secure from physical attacks such as differential power analysis attacks \cite{kocher1999differential} and timing attacks \cite{kocher1996timing}. Given a specific cache model and a specific prediction model, we assume an adversary who can control the provided inputs to the program and perform cache timing attacks via another thread or process on the same device at any time during execution. Commonly, the inputs provided to the program may correspond to user input, a specific data received via network, or a pre-defined input. We consider common cache timing attacks such as FLUSH+RELOAD \cite{yarom2014flush, gruss2016flush} or PRIME+PROBE \cite{osvik2006cache, percival2005cache, bonneau2006cache, liu2015last, tromer2010efficient} where the adversary can measure the access time for each cache line or cache set. In our model, branch predictors are process-specific, so we only consider same address-space training \cite{canella2019systematic}. However, the adversary is able to make the victim process use full length of speculation window by mounting denial-of-service (DoS) attacks \cite{woo2007analyzing, bechtel2019denial} which can block shared hardware resources that are necessary for evaluation of the prediction outcome.

\section{Prediction-aware Symbolic Execution}
\label{sec:predictionaware}
Specognitor's 4 major units as presented in Figure~\ref{img:Specognitor_Architecture} are the symbolic execution engine, the speculative execution model, the cache model, and the monitor. Specognitor also combines taint analysis with symbolic execution to identify secret-dependent values. In the following we elaborate on each unit.

\subsection{Symbolic Execution Engine}
\begin{figure}[t]
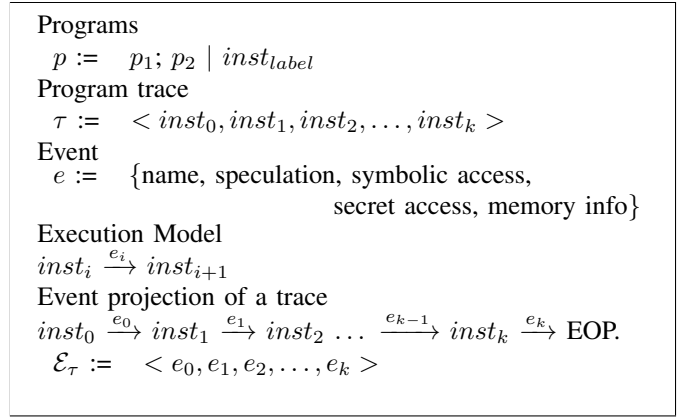

\begin{mdframed}
Programs \\
\begin{tabular}{ll}
$p$ := & $p_1$; $p_2$ $\mid$ $inst_{label}$ \\
\end{tabular} \\
Program trace \\
\begin{tabular}{ll}
$\tau$ := & $<inst_0, inst_1, inst_2, \dots, inst_k>$ \\
\end{tabular} \\
Event \\
\begin{tabular}{ll}
$e$ := & \{name, speculation, symbolic access, \\
        & \hspace{26mm} secret access, memory info\}
\end{tabular} \\
Execution Model \\
$inst_i$ $\xrightarrow[]{e_i}$ $inst_{i+1}$ \\
Event projection of a trace \\
$inst_0$ $\xrightarrow[]{e_0}$ $inst_1$ $\xrightarrow[]{e_1}$ $inst_2$ $\dots$  $\xrightarrow[]{e_{k-1}}$ $inst_k$ $\xrightarrow[]{e_k}$ EOP.  \\
\begin{tabular}{l l}
$\mathcal{E}_\tau$ := & $<e_0, e_1, e_2, \dots, e_k>$ \\
\end{tabular} \\
\end{mdframed}
\caption{Program and execution semantics.}
\label{img:semantics}
\end{figure}

Specognitor's symbolic execution engine takes a program $\mathcal{P}$ that is a sequence of LLVM \cite{llvmReferenceManual, lattner2004llvm} instructions and symbolically executes it w.r.t. the speculative execution model. Symbolic engine generates program traces according to the input values and the CFG. 
According to Figure~\ref{img:semantics}, a program trace $\tau$ is a sequence of instructions determined by the CFG and symbolic values. We use $\mathcal{T}_\mathcal{P}$ to denote the set of all traces of program $\mathcal{P}$. 
A program trace represents a complete execution of the program w.r.t. a specific set of inputs and the path condition ($\psi$). $\tau_\psi$ is a program trace in which the path condition $\psi$ holds. For example, in Figure~\ref{img:Example1}, the two program traces are $\tau_{(x \geq array1\_{}size)}$ and $\tau_{(x < array1\_{}size)}$. Execution of the program trace $\tau$ results in generating a sequence of events $\mathcal{E}_\tau$. These events are the communication channel between the execution engine and other components such as prediction logic and cache. Through these events a processor can modify the micro-architectural state of the components. Events are trace-specific that is execution of identical instructions in two different traces might generate different events. An event convey multiple information about the instruction: i) name, ii) the speculation status, iii) whether it uses symbolic operands, iv) whether it uses secret-dependent operands, and v) addresses accessed by memory access instructions (i.e. Store and Load).

\subsection{Speculative Execution Model}
We parameterize our speculative model on two main features:

\noindent $\bullet$ Prediction model: we model two types of branch predictors. The two-level prediction logic is only used for conditional branches and parameterized by the size of BHR. For example, PHT4 corresponds to a two-level branch prediction where the first level uses a 4-bit BHR and the second level uses a $2^4$-entry PHT. The branch target buffer can be used for both conditional branches and indirect branches. It is parameterized by the number of BTB sets, the number of BTB ways, and the number of tag bits.

\noindent $\bullet$ Speculative execution window ($\omega$): in out-of-order processors, instructions are executed speculatively from the reorder buffer, so the size of the reorder buffer is considered as the maximum depth of speculative execution. On the other hand, in in-order pipeline processors, the number of pipeline stages represents an upper bound for the speculative execution window \cite{doweck2017inside}.

Based on these two features, Specognitor generates full-length speculative traces during execution w.r.t. $\omega$ on conditional branches and indirect branches/calls. In Figure~\ref{img:Example1}, for the $\tau_{(x \geq array1\_{}size)}$ program trace, when the symbolic execution engine reaches the branch instruction, it asks the prediction logic for next instruction. If the predictor mispredicts the outcome and returns line 5 as the next instruction, Specognitor modifies the program trace and inserts the speculative trace. Since there might be other branch instructions in the speculative trace, Specognitor applies symbolic execution from the new mispredicted instruction and generate all possible speculative traces w.r.t. the prediction logic up to the length of $\omega$. Spcognitor uses the prediction logic for all speculative and non-speculative branches. When the speculative traces are executed, Specognitor updates the branch predictor only for the non-speculative branch instruction, that is line 4 in Figure~\ref{img:Example1}.


\subsection{Cache Component}
To model the cache behavior, we use the same cache modeling as \textsc{KLEESpectre} \cite{wang2020kleespectre}. In this model, the cache captures memory access operations and tries to model the cache behavior, i.e., loading a cache line that might evict previous cache lines. Since our cache modeling follows prior work \cite{wang2020kleespectre}, we eliminate the model description. However, we define a communication event $\mathcal{C}$ between cache and the monitor. Through this communication channel the cache component can send information about its state to the monitor. $\mathcal{C}_{i}$ shows the state of the cache after receiving the event $e_i$.

\subsection{Pattern Detection Monitor}
To keep records of the system state during execution, we propose a monitoring component. We present our monitor as a directed path, i.e., a finite sequence of distinct nodes which are connected by directed edges (see Figure~\ref{img:MonitorDesc}). While executing a program trace, our monitor tries to fire transitions and propagate tokens. A token is a 4-tuple $t=\langle id, pid, inst, ttl \rangle$ where $id$ is the unique identifier of the token, $pid$ is the id of its predecessor, $inst$ is a structure containing instruction information, and $ttl$ is the liveness of the token ($-1$ if it is not defined). The monitor copies and moves a token from $Node_i$ to $Node_{i+1}$ when it observes events ($e_{i+1}$,$\mathcal{C}_{i+1}$) that satisfies all properties specified by $Node_{i+1}$ ($np_{i+1}$). Obviously, $Node_i$ should contain a token before propagating to $Node_{i+1}$. The transition rule of a token can be formalized as follows:

\begin{figure}
\centering
\begin{mdframed}
A monitor is a 5-tuple ($Q$, $\Sigma$, $\delta$, $q_0$, $F$), where: \\
\begin{tabular}{ll}
$Q$ := & $\{Start, Node_1, Node_2, ..., Node_n\}$ \\
$\Sigma$ := & $\{np_1, np_2, ..., np_n\}$ \\
$\delta$ := & $\Bigl\{ \langle Start, Node_1 \rangle, \langle Node_1, Node_2 \rangle, ...$ \\ 
& \hfill $,\langle Node_{n-1}, Node_n \rangle \Bigl\}$ \\
$q_0$ := & $Start$ \\
$F$ := & $\{Node_n\}$ \\
\end{tabular}
\end{mdframed}
\caption{\small Formal description of a monitor.}
\label{img:MonitorDesc}
\end{figure}

\begin{figure}
\centering
\includegraphics[width=0.35\textwidth, height=0.18\textheight]{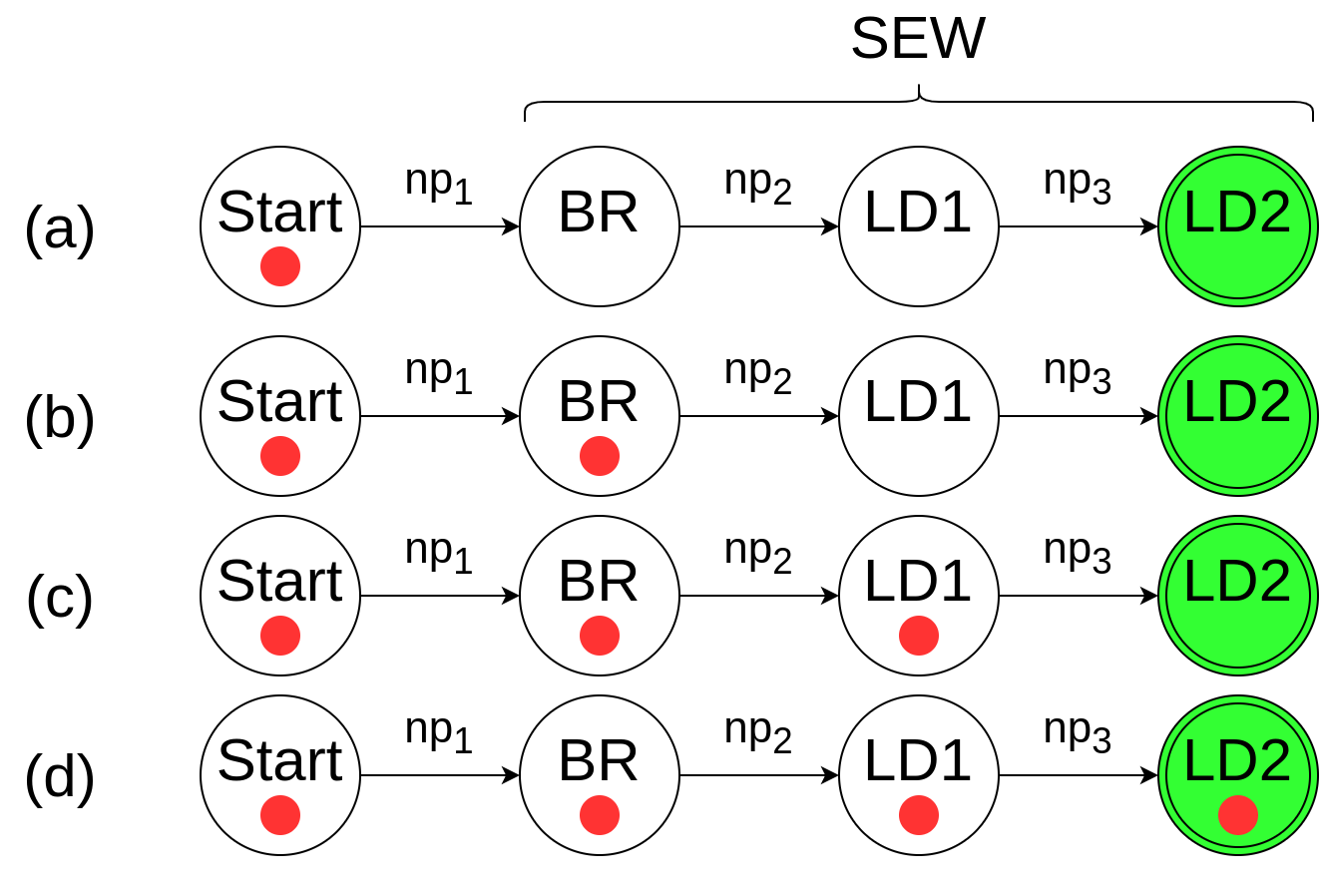}
\caption{\small Different states of Specognitor's pattern detection mechanism for {\ttfamily BR-LD-LD} pattern description. (a) shows the state of pattern detector in the beginning of execution. (b) shows the state of the pattern detector after detecting a node that satisfies $np_1$. (c) shows the state of the pattern detector after detecting a node that satisfies $np_2$. (d) shows the state of the pattern detector after detecting a node that satisfies $np_3$.}
\label{img:PatternFSM}
\end{figure}

\begin{equation}
\begin{split}
    (e_j,\mathcal{C}_{j} \models np_{i+1}) \; \& \; (\langle id, pid, inst_k, ttl \rangle \in Node_i) \\ 
    \Rightarrow (\langle id', id, inst_j, ttl \rangle \in Node_{i+1})
\end{split}
\label{eq:transition}
\end{equation}

The monitor's node properties ($np$) are as follows:

\noindent $\bullet$ \textit{Instruction:} specifies the name of instruction such as branch, call, load, store, and etc. \\
$\bullet$ \textit{isSpeculative:} checks whether the instruction is executed in the wrong speculative instruction stream w.r.t. the prediction logic. \\
$\bullet$ \textit{isConst:} checks whether the operands are symbolic. \\
$\bullet$ \textit{isSensitive:} checks whether the operands are secret-dependent. \\
$\bullet$ \textit{checkCacheState:} after executing the instruction, checks whether the cache is vulnerable against cache attacks. \\
$\bullet$ \textit{startTTL:} starts a time-to-live counter for the token and all its successors. This property is used for checking whether certain instructions are executed within a window. \\
$\bullet$ \textit{stopTTL:} stops the time-to-live counter. This property specifies the end of the execution counter for instructions.

The initial node of the monitor is always the $Start$ node in which there is a token. Along execution, the execution engine sends events to the monitor. When a token reaches a node with \textit{startTTL} property, the monitor starts a counter for it. It decrements the counter upon receiving an event. If the token reaches a node in which \textit{stopTTL} property is true, the monitor stops the counter. Otherwise, the monitor decrements the counter until it reaches $0$ and expires the token.

We use the following definitions to describe a leakage-free program.

\begin{definition}
\label{def:def1}
In an execution trace, if there is a pair of events $(e_k,\mathcal{C}_k)$ that satisfies $np_i$, then $P_{(Node_i)}$ is true in the execution trace. In other words, $P_{(Node_i)}$ is true iff $(e_k,\mathcal{C}_{k}) \models np_{i}$.
\end{definition}

\begin{definition}
\label{def:def2}
$P_{(Node_i \rightarrow Node_{i+1} \dots \rightarrow Node_j)}$ is true in an execution trace, iff there exist a pair of events $(e_k,\mathcal{C}_k)$ that satisfies $np_j$ while there is at least one path from $i$ to $j-1$ in which all nodes contain at least one token.
\end{definition}

According to definition ~\ref{def:def1} and ~\ref{def:def2}, we express a leakage free program as follows:

\begin{equation}
\begin{split}
\nexists \; \tau \in \mathcal{T}_\mathcal{P} : \; & P_{(Node_1)} \; \& \\ 
& P_{(Node_1 \rightarrow Node_2)} \; \& \\ 
& P_{(Node_1 \rightarrow Node_2 \rightarrow Node_3)} \; \& \\
& \dots \; \& \\ 
& P_{(Node_1 \rightarrow Node_2 \rightarrow \dots \rightarrow Node_n)}
\end{split}
\label{eq:leakageFree}
\end{equation}

A program $\mathcal{P}$ is leakage-free w.r.t. the micro-architectural configurations and pattern model if there is no trace $\tau$ in which the monitor can propagate a token to the last node. Figure~\ref{img:PatternFSM} shows the monitor workflow in detecting the leakage pattern for the code snippet represented in Figure~\ref{img:01}. In the beginning, the pattern detector starts from the $Start$ node. When the symbolic execution engine reaches the branch instruction at line 4, it generates an event which satisfies $np_1$. Therefore, the monitor propagates a token to node $BR$. Then, the execution engine speculatively executes the load instruction at line 5 which loads the secret. The events generated from this execution satisfy $np_2$ and the monitor propagates the token to the next node. Finally, the execution engine speculatively executes the load instruction at line 6 that leaks the secret and generates $e_6$. After loading a secret-dependent address to cache, the cache model generates $\mathcal{C}_6$ which indicates that the cache is vulnerable to cache attacks. The generated events $e_6$ and $\mathcal{C}_6$ satisfy $np_3$, so the monitor propagates the token to the last node ($Node_3$) and reports a new leakage pattern. Checking the cache state is optional and can be used according to the value of \textit{checkCacheState} defined by the user.

\subsection{Taint Analysis}
In Specognitor, we adopt taint analysis to propagate the dependency of instructions and memory locations on secret keys. Since LLVM instruction set uses virtual registers in Static Single Assignment \cite{cytron1991efficiently} form, it is crucial to propagate the secrecy of key values so that the generated events in the subsequent instructions can use it for detecting leakage. For example, in code fragment of Figure~\ref{img:01}, after loading the secret value of {\ttfamily array1[x]} at line 5 and storing it in the {\ttfamily temp} variable, it is important to taint the {\ttfamily temp} variable so that we detect secret-dependent memory access at line 6 ({\ttfamily array2[temp*64]}). We summarize the taint propagation rules in Table~\ref{tab:taintRules}. According to Table~\ref{tab:taintRules}, the result of binary and unary operations is tainted if the operand(s) are tainted. For $store$ instructions, the content of the memory location with address $memLoc$ is tainted if the content of register $reg$ is tainted. Similarly, for $load$ instructions, the content of resigter $reg$ is tainted if the content of memory location with address $memLoc$ is tainted.

\begin{table}
\caption{\small Taint propagation rules. $\ominus$ represents a unary operator. $\otimes$ represents a binary operator.}
\begin{center}
\begin{tabular}{|l|l|}
\hline
\textbf{Expression} & \textbf{Description} \\
\hline
$e_2 = \ominus e_1$ & $e_2$ is tainted if $e_1$ is tainted. \\
\hline
$e_3 = e_1 \otimes e_2$ & $e_3$ is tainted if $e_1$ or $e_2$ are tainted. \\
\hline
$\textbf{store}$ $memLoc,$ $reg$ & \makecell[l]{The memory content indexed by $memLoc$ \\ is tainted if the content of $reg$ is tainted.} \\
\hline
$\textbf{load}$ $reg,$ $memLoc$ & \makecell[l]{The content of $reg$ is tainted if the memory \\ content indexed by $memLoc$ is tainted.} \\
\hline
\end{tabular}
\label{tab:taintRules}
\end{center}
\end{table}

\section{Implementation}
\label{sec:implementation}
\begin{figure}
\centering
\includegraphics[scale=0.15]{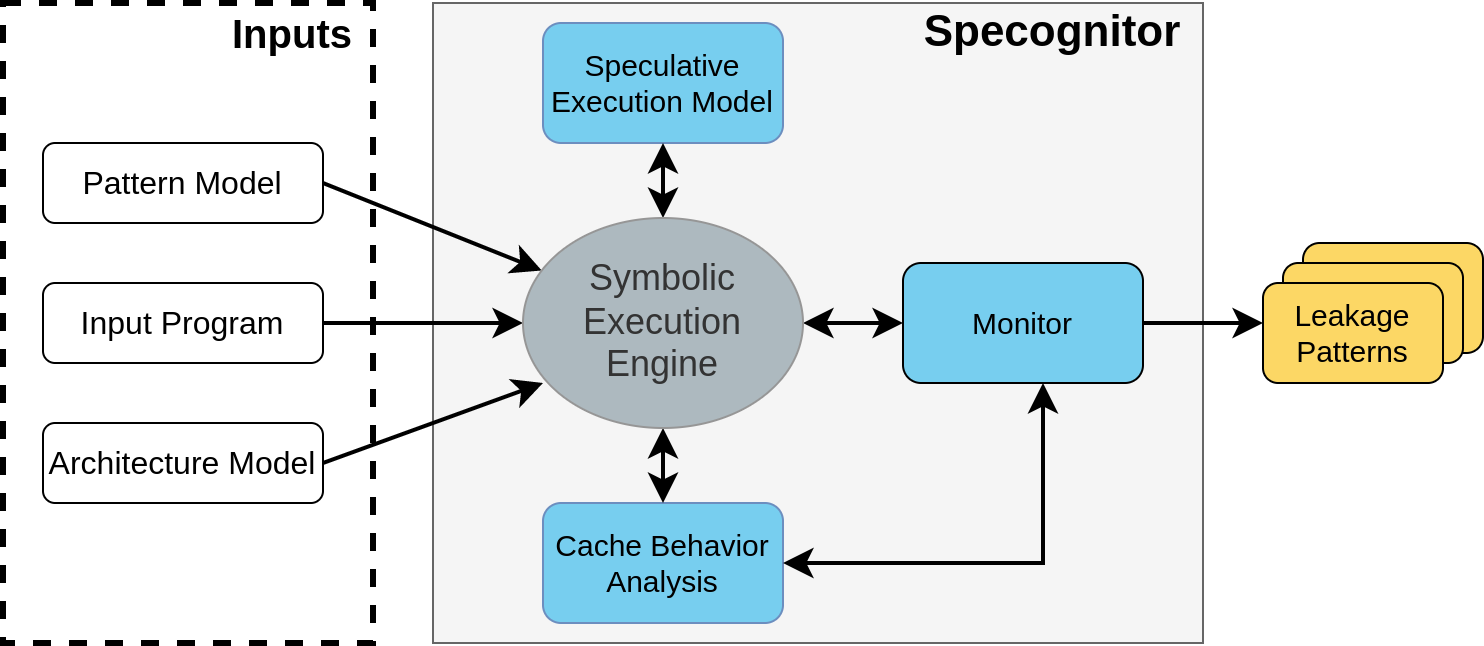}
\caption{\small Front-end view of Specognitor.}
\label{img:Specognitor_Architecture}
\end{figure}

We developed Specognitor on top of the state-of-the-art symbolic execution engine KLEE v2.1 \cite{cadar2008klee}. It also adopts cache state modeling from KLEESpectre \cite{wang2020kleespectre}. Specognitor uses clang v6.0 as the back-end compiler and takes programs in LLVM bitcode format generated with LLVM-6.0. In addition, it takes pattern models in JSON format and architecture model as command line options. Input programs containing external function calls are linked with KLEE-uClibc \cite{ucLib} before being processed by Specognitor. We use the SMT solver STP \cite{ganesh2007decision} to check the satisfiability of symbolic formulas during execution. Figure~\ref{img:Specognitor_Architecture} shows the front-end view of Specognitor.


\section{Evaluation Results}
\label{sec:evaluation}
In this section, we present our evaluation results on effectiveness, real-world applicability, and scalability of the Specognitor. We evaluate the tool by answering the following research questions:

\begin{itemize}
\item[RQ1:] How effective is Specognitor in detecting different types of Spectre variant 1 and Spectre variant 2?
\item[RQ2:] How effective is the architecture modeling of Specognitor w.r.t. real-world processors?
\item[RQ3:] Can Specognitor detect vulnerabilities in real-world cryptographic programs?
\item[RQ4:] How computationally efficient is the prediction-aware symbolic execution compared to a state-of-the-art tool without prediction logic?
\item[RQ5:] How computationally expensive is the prediction-aware symbolic execution compared to the pure symbolic execution?
\end{itemize}

\textbf{Experimental Setup.} To conduct our experimental evaluation, we run our benchmarks on a quad-core Intel i7-8665U processor with 16GB of RAM on Ubuntu 18.04.6 LTS. we compile our programs using Clang-6.0 with \textit{-O0} optimization level. Since cache modeling is not part of the main contributions of this work and mostly follows prior work \cite{wang2020kleespectre}, we do not consider it in our evaluation results. All BTB configurations are direct-mapped and parameterized by the number of sets unless otherwise stated.

\subsection{Pattern Detection}
In the first step, we focus on testing effectiveness of Specognitor on detecting different vulnerability patterns (RQ1) w.r.t. spectre variant 1 and spectre variant 2.

\begin{figure}
  \begin{lstlisting}[language=C,numbers=left,stepnumber=1,frame=lines,basicstyle=\ttfamily\scriptsize,xleftmargin=3em,framexleftmargin=2em]
uint8_t victim_fun(int idx)
{
    uint8_t temp = 0;
    if (idx >= array1_size) {
	    /* dummy computation */
    } else {
	    temp = array1[idx];
	    temp &= array2[temp*64];
    }
    return temp;
}
  \end{lstlisting}
  \caption{\small Example v02 (complement of example v01)}
  \label{img:v02}
\end{figure}

Initially, we use two sets of litmus tests: first, we take 15 examples proposed by Kocher et al. \cite{kocher2020spectre} to perform spectre variant 1 attacks targeting PHT of the processor. We add an examples to these 15 examples to create a benchmark of 16 PHT test cases. We add example v02 (Figure~\ref{img:v02}) as a complement of example v01 (Figure~\ref{img:01}) to show the effectiveness of Specognitor compared to the state-of-the-art tools. In both examples, line 4 is the start of speculation and we assume that the state of the prediction logic is identical at line 4. By accounting for the prediction logic, only one of them should be treated as vulnerability and the other one is safe. Second, according to the 16 PHT test cases, we created 16 test cases to target BTB of the processors to address architectures in which two-level prediction logic is not used (e.g. Pentium 4 \cite{hinton2001microarchitecture}). We evaluate Specognitor on these 32 examples with different prediction parameters and speculative execution windows. To keep the code as small as possible, we do not train the branch predictors. Instead, we used very simplified versions of PHT and BTB with 1-4 BHR-bits and 1-4 BTB entries and 16-32 speculative execution window to show the effect of different architecture models on information leakage. We test 32 example on 8 different configurations (total of 256 litmus examples) and Specognitor could successfully detect all vulnerabilities as intended. Table~\ref{tab:pht_examples} and~\ref{tab:btb_examples} show the vulnerabilities detected in each code example for different configurations. We highlight the following findings:

\noindent $\bullet$ In example v01 and v02 from PHT benchmark, we can see that Specognitor only detects leakage pattern in one of them while recent approaches \cite{wang2020kleespectre, brotzman2021specsafe, guarnieri2020spectector} mark both examples as vulnerable. \\
$\bullet$ In PHT benchmarks, we mainly use {\ttfamily BR-LD-LD} pattern, except for example v11 and v12 where the leakage is caused by branching on key-dependent value. For these examples, we use a {\ttfamily LD-BR-LD} pattern, where i) the first {\ttfamily LD} loads a secret key, ii) the {\ttfamily BR} instruction uses the secret value in a comparison, and iii) the last node is a simple load instruction. 
\begin{lstlisting}[language=C,numbers=left,stepnumber=1,frame=lines,basicstyle=\ttfamily\footnotesize,xleftmargin=3em,framexleftmargin=2em]
key = array[idx];
if(key == k){ temp &= array[0]; }
\end{lstlisting}
The adversary can infer key value by checking whether {\ttfamily array[0]} is loaded in the cache. By using this pattern description, Specognitor could also successfully detect the three speculative side-channels introduced by \cite{brotzman2021specsafe}. These attacks are all based on a key-dependent branch instruction.

\begin{table}
\centering
\caption{\small PHT examples. ($\bullet$) indicates a leakage of {\ttfamily BR-LD-LD} pattern, ($\star$) indicates a leakage of {\ttfamily LD-BR-LD} pattern, and ($\circ$) indicates no leakage.}
{\tiny
\begin{tabular}{| c | c c | c c | c c | c c |}
\hline
PHT & \multicolumn{2}{c |}{$\omega=16$} & \multicolumn{2}{c |}{$\omega=32$} & \multicolumn{2}{c |}{$\omega=16$} & \multicolumn{2}{c |}{$\omega=32$} \\
Benchmarks & PHT:1 & PHT:4 & PHT:1 & PHT:4 & BTB:1 & BTB:4 & BTB:1 & BTB:4 \\
\hline
v01 & $\bullet$ & $\bullet$ & $\bullet$ & $\bullet$ & $\bullet$ & $\bullet$ & $\bullet$ & $\bullet$ \\
v02 & $\circ$ & $\circ$ & $\circ$ & $\circ$ & $\circ$ & $\circ$ & $\circ$ & $\circ$ \\
v03 & $\bullet$ & $\bullet$ & $\bullet$ & $\bullet$ & $\bullet$ & $\bullet$ & $\bullet$ & $\bullet$ \\
v04 & $\bullet$ & $\bullet$ & $\bullet$ & $\bullet$ & $\bullet$ & $\bullet$ & $\bullet$ & $\bullet$ \\
v05 & $\bullet$ & $\bullet$ & $\bullet$ & $\bullet$ & $\bullet$ & $\bullet$ & $\bullet$ & $\bullet$ \\
v06 & $\bullet$ & $\bullet$ & $\bullet$ & $\bullet$ & $\bullet$ & $\bullet$ & $\bullet$ & $\bullet$ \\
v07 & $\bullet$ & $\bullet$ & $\bullet$ & $\bullet$ & $\bullet$ & $\bullet$ & $\bullet$ & $\bullet$ \\
v08 & $\bullet$ & $\bullet$ & $\bullet$ & $\bullet$ & $\bullet$ & $\bullet$ & $\bullet$ & $\bullet$ \\
v09 & $\bullet$ & $\bullet$ & $\bullet$ & $\bullet$ & $\bullet$ & $\bullet$ & $\bullet$ & $\bullet$ \\
v10 & $\bullet$ & $\bullet$ & $\bullet$ & $\bullet$ & $\bullet$ & $\bullet$ & $\bullet$ & $\bullet$ \\
v11 & $\star$ & $\star$ & $\star$ & $\star$ & $\star$ & $\star$ & $\star$ & $\star$ \\
v12 & $\star$ & $\star$ & $\star$ & $\star$ & $\circ$ & $\star$ & $\star$ & $\star$ \\
v13 & $\bullet$ & $\bullet$ & $\bullet$ & $\bullet$ & $\bullet$ & $\bullet$ & $\bullet$ & $\bullet$ \\
v14 & $\bullet$ & $ \bullet$ & $\bullet$ & $\bullet$ & $\bullet$ & $\bullet$ & $\bullet$ & $\bullet$ \\
v15 & $\bullet$ & $\bullet$ & $\bullet$ & $\bullet$ & $\bullet$ & $\bullet$ & $\bullet$ & $\bullet$ \\
v16 & $\bullet$ & $\bullet$ & $\bullet$ & $\bullet$ & $\bullet$ & $\bullet$ & $\bullet$ & $\bullet$ \\
\hline
\end{tabular}}
\label{tab:pht_examples}
\end{table}

\noindent $\bullet$ In BTB benchmark, we observe no leakage for PHT-based configurations except for example v09. However, in this example, the leakage detected for PHT-based configurations and BTB-based configurations happen in two different locations.
\begin{lstlisting}[language=C,numbers=left,stepnumber=1,frame=lines,basicstyle=\ttfamily\footnotesize,xleftmargin=2em,framexleftmargin=2em]
uint32_t x = 0;
temp&=array2[array1[x<(array1_size-1)?(x+1):0]];
x = idx;
temp&=array2[array1[x<(array1_size-1)?(x+1):0]];
\end{lstlisting}
In this code example, for configurations with 1-entry BTB, the leakage happens at line 2 while for other configurations (4-entry BTB and all PHT-based configurations) leakage happens at line 4. The leakage pattern detected at line 4 corresponds to the normal {\ttfamily BR-LD-LD}. However, the leakage pattern at line 2 is more complex and corresponds to the training of the prediction logic. To expose this leakage, the execution engine executes the instruction at line 2 and since the condition {\ttfamily x<(array1\_{}size-1)} is true, it takes {\ttfamily x+1} as index. By executing this line, the single BTB entry is updated. At line 3, the value of {\ttfamily x} is changed to a symbolic value ({\ttfamily idx}). When the execution engine reaches the condition at line 4, it speculatively jumps to the instruction at line 2 and executes it with {\ttfamily x+1} index which is symbolic now and might cause information leakage. This example illustrates the importance of using a precise prediction logic and pattern descriptions to detect the root cause of the vulnerability.

\begin{table}
\centering
\caption{\small BTB examples. ($\bullet$) indicates a leakage of {\ttfamily BR-LD-LD} pattern, ($\star$) indicates a leakage of {\ttfamily LD-BR-LD} pattern, and ($\circ$) indicates no leakage.}
{\tiny
\begin{tabular}{| c | c c | c c | c c | c c |}
\hline
BTB & \multicolumn{2}{c |}{$\omega=16$} & \multicolumn{2}{c |}{$\omega=32$} & \multicolumn{2}{c |}{$\omega=16$} & \multicolumn{2}{c |}{$\omega=32$} \\
Benchmarks & PHT:1 & PHT:4 & PHT:1 & PHT:4 & BTB:1 & BTB:4 & BTB:1 & BTB:4 \\
\hline
v01 & $\circ$ & $\circ$ & $\circ$ & $\circ$ & $\bullet$ & $\circ$ & $\bullet$ & $\circ$ \\
v02 & $\circ$ & $\circ$ & $\circ$ & $\circ$ & $\bullet$ & $\circ$ & $\bullet$ & $\circ$ \\
v03 & $\circ$ & $\circ$ & $\circ$ & $\circ$ & $\circ$ & $\circ$ & $\circ$ & $\circ$ \\
v04 & $\circ$ & $\circ$ & $\circ$ & $\circ$ & $\circ$ & $\bullet$ & $\circ$ & $\bullet$ \\
v05 & $\circ$ & $\circ$ & $\circ$ & $\circ$ & $\bullet$ & $\circ$ & $\bullet$ & $\circ$ \\
v06 & $\circ$ & $\circ$ & $\circ$ & $\circ$ & $\circ$ & $\circ$ & $\circ$ & $\circ$ \\
v07 & $\circ$ & $\circ$ & $\circ$ & $\circ$ & $\bullet$ & $\circ$ & $\bullet$ & $\circ$ \\
v08 & $\circ$ & $\circ$ & $\circ$ & $\circ$ & $\circ$ & $\circ$ & $\circ$ & $\circ$ \\
v09 & $\bullet$ & $\bullet$ & $\bullet$ & $\bullet$ & $\bullet$ & $\bullet$ & $\bullet$ & $\bullet$ \\
v10 & $\circ$ & $\circ$ & $\circ$ & $\circ$ & $\circ$ & $\circ$ & $\circ$ & $\circ$ \\
v11 & $\circ$ & $\circ$ & $\circ$ & $\circ$ & $\circ$ & $\circ$ & $\star$ & $\circ$ \\
v12 & $\circ$ & $\circ$ & $\circ$ & $\circ$ & $\circ$ & $\circ$ & $\circ$ & $\circ$ \\
v13 & $\circ$ & $\circ$ & $\circ$ & $\circ$ & $\bullet$ & $\circ$ & $\bullet$ & $\circ$ \\
v14 & $\circ$ & $\circ$ & $\circ$ & $\circ$ & $\bullet$ & $\circ$ & $\bullet$ & $\circ$ \\
v15 & $\circ$ & $\circ$ & $\circ$ & $\circ$ & $\circ$ & $\circ$ & $\circ$ & $\circ$ \\
v16 & $\circ$ & $\circ$ & $\circ$ & $\circ$ & $\bullet$ & $\circ$ & $\bullet$ & $\circ$ \\
\hline
\end{tabular}}
\label{tab:btb_examples}
\end{table}

Moreover, to address spectre variant 2, we build a small program according to the BTB attacks mounted by Canella et al. \cite{canella2019systematic} to evaluate the effectiveness of Specognitor in detecting Spectre variant 2 attacks. Figure~\ref{img:03} represents the idea behind the example designed for Spectre variant 2. We run Specognitor with different configurations to see whether it can detect the information leakage. Table~\ref{tab:spectrev2} shows the results of our experiment. As can be seen from the Table~\ref{tab:spectrev2}, configurations without BTB are not vulnerable to this variant of spectre attacks. Furthermore, in 1-entry BTB configurations (without PHT), the malicious BTB entry is evicted, therefore these configurations do not expose any information leakage. However, other configurations report vulnerability at line 7 in the code presented in Figure~\ref{img:03}.

\begin{table}
\centering
\caption{\small Spectre-v2 vulnerability detection. Symbols show vulnerable ($\bullet$) and safe ($\circ$) configurations.}
{\tiny\sffamily
\setlength\tabcolsep{3.5pt}
\begin{tabular}{| c | c  c | c  c | c  c | c  c | c | c |}
\hline
\multirow{2}{*}{Config.} & \multicolumn{2}{c |}{$\omega=16$} & \multicolumn{2}{c |}{$\omega=32$} & \multicolumn{2}{c |}{$\omega=16$} & \multicolumn{2}{c |}{$\omega=32$} & $\omega=16$ & $\omega=32$ \\
 & PHT:1 & PHT:4 & PHT:1 & PHT:4 & BTB:1 & BTB:4 & BTB:1 & BTB:4 & \makecell{PHT:4 \\ BTB:4} & \makecell{PHT:4 \\ BTB:4} \\
\hline
Leakage & $\circ$ & $\circ$ & $\circ$ & $\circ$ & $\circ$ & $\bullet$ & $\circ$ & $\bullet$ & $\bullet$ & $\bullet$ \\
\hline
\end{tabular}}
\label{tab:spectrev2}
\end{table}

\subsection{Real World Experiments}

\begin{table}
\centering
\caption{\small Vulnerability detection on Libgcrypt with real processor configurations.}
{\tiny\sffamily
\begin{tabular}{| c | c | c | c |}
\hline
\multirow{2}{*}{Config.} & ARM - Cortex-A53 & ARM - Cortex-A7 & Pentium 4 \\
 & $\omega=32$,PHT:12-4096,BTB:256 & $\omega=16$,PHT:8-256,BTB:8 & $\omega=40$,BTB:4096\\
\hline
str2key & $60$ & $30$ & $58$ \\
\hline
\end{tabular}}
\label{tab:realProc}
\end{table}

In the next step, we designed an experiment in which Specognitor executes the real-world cryptographic programs from libgcrypt library such as {\ttfamily  hpn-ssh}, {\ttfamily openssl}, {\ttfamily Linux-tegra}, and {\ttfamily libTomCrypt} w.r.t. 3 real world processors (Cortex-A53, Cortex-A7, Pentium 4) to address RQ2 and RQ3. We highlight the following results:

\noindent $\bullet$ Specognitor could not find any memory access in which the address depends on a secret key in speculative traces except for \textit{str2key} program. Line 2 of the following code fragment shows the vulnerable code segment of {\ttfamily str2key} program:
\begin{lstlisting}[language=C,numbers=left,stepnumber=1,frame=lines,basicstyle=\ttfamily\footnotesize,xleftmargin=3em,framexleftmargin=2em]
for (i=0; i<DES_KEY_SZ; i++)
    (*key)[i]=odd_parity[(*key)[i]];
\end{lstlisting}
Our results show that Specognitor reports vulnerability w.r.t. the three processors while executing this code fragment. However, based on the architectural specifications of a processor, the repetitions of line 2 in speculative paths are different. Table~\ref{tab:realProc} shows the number of detected vulnerabilities according to the configuration.


\noindent $\bullet$ According to Table~\ref{tab:realProc}, the size of speculation window $\omega$ and the prediction logic both affect the number of vulnerabilities detected on the configurations.

\subsection{Scalibility}
In this section, we compare the computational cost of Specognitor with \textsc{KLEESpectre} \cite{wang2020kleespectre} and KLEE \cite{cadar2008klee}. To this end, we perform two sets of experiments. For all experiments in this section, we use 10 real-world cryptographic programs from libgcrypt library as the evaluation data set.

\begin{figure*}[!b]
    \centering
    \includegraphics[width=\textwidth, height=0.25\textheight]{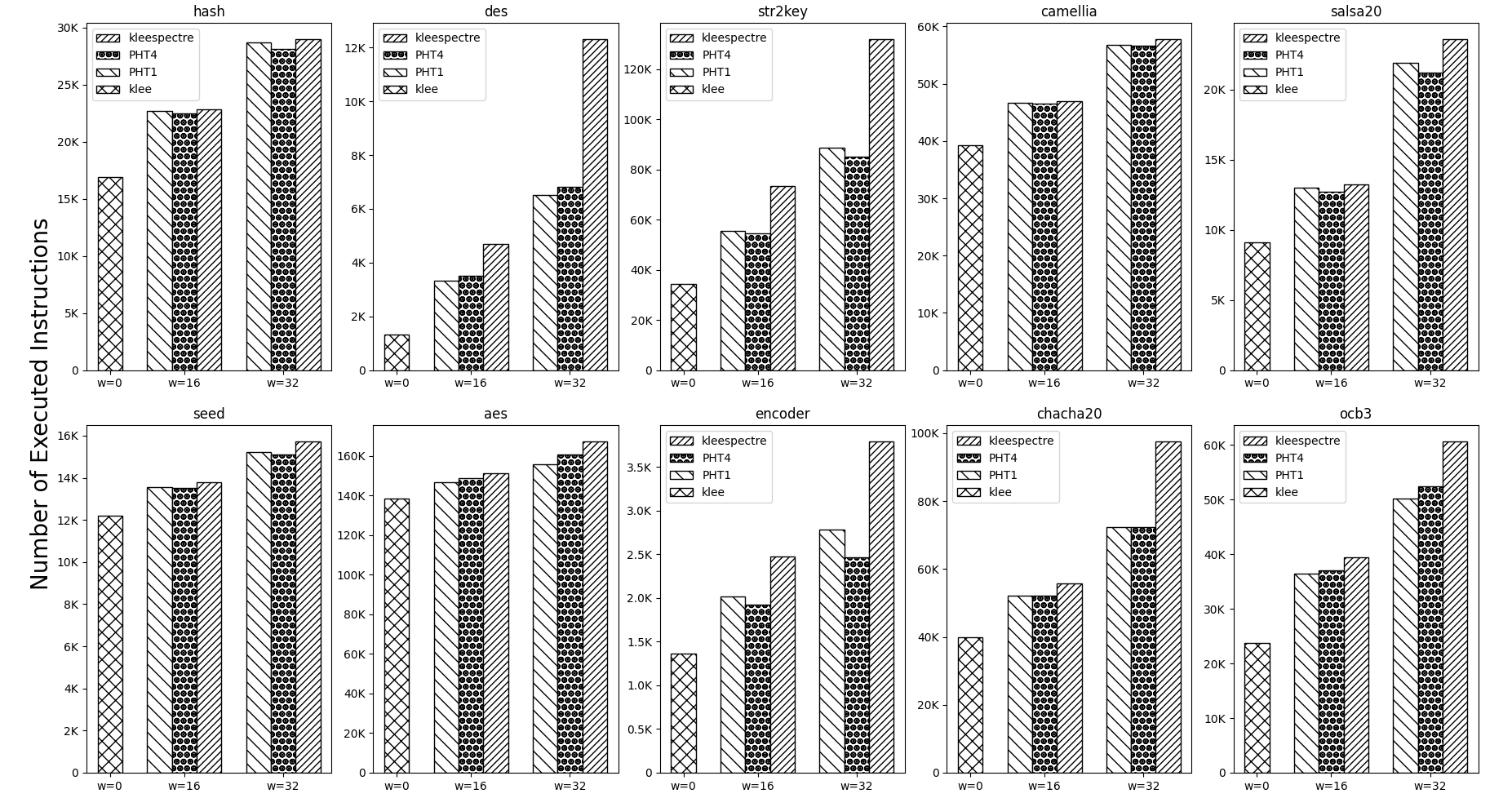}
    \caption{\small The number of executed instructions for different Libgcrypt programs on klee, kleespectre, and two different prediction configuration of specognitor, i.e., PHT1, and PHT4. $\omega=0$ presents experimental results of klee. $\omega=16$ and $\omega=32$ presents experimental results of kleespectre and specognitor (PHT1 and PHT4).}
    \label{fig:instructionCount}
\end{figure*}

\noindent $\bullet$ First, to show the efficiency of our tool and its ability to prune infeasible paths (RQ4), we compare the test results to \textsc{KLEESpectre} \cite{wang2020kleespectre}, an state-of-the-art speculation-aware symbolic execution engine developed on top of KLEE, but lacks the prediction modeling capability and dynamic pattern detection. To have a meaningful comparison, we use PHT-based configurations of Specognitor since \textsc{KLEESpectre} does not implement BTB. We execute each program on KLEE, on \textsc{KLEESpectre} with two different speculative execution length (16-32) and on Specognitor with four different configurations, i.e., 1-bit and 4-bit two level branch prediction and two different speculative exeution length (16-32). Figure~\ref{fig:instructionCount} visualize the comparison of the total amount of executed instructions. The experimental results are shown in three groups: i) without speculation, ii) $\omega=16$, and iii) $\omega=32$. As expected, Specognitor removes the infeasible speculative instruction traces. According to the results, Specognitor shrinks the execution space by up to $\sim\!48\%$ for {\ttfamily DES} program (on average by $\sim\!15\%$ for all programs).

\noindent $\bullet$ Second, we compare Specognitor with KLEE symbolic execution engine to assess the computation cost of prediction modeling and symbolic execution (RQ5). Since Specognitor is developed on top of KLEE, the experimental results give us the exact amount of computation cost added to the symbolic execution due to the speculative path exploration and prediction logic. As expected, the speculation modeling of the Specognitor increases the execution time. According to the results shown in Table~\ref{tab:exeTime}, execution time of programs on Specognitor on average takes $\sim\!10$x more than their execution time on KLEE, for $\omega=16$ ($\sim\!20$x for $\omega=32$). However, for 7 out of 10 programs, it takes less than 1 second to execute on KLEE. For examples which require more than 1 second to execute on KLEE ({\ttfamily str2key, camellia,} and {\ttfamily seed}), the execution time of Specognitor is on average $\sim\!17\%$ more than KLEE, for $\omega=16$ ($\sim\!22\%$ for $\omega=32$). This illustrates that Specognitor has a static computation cost due to the monitor and prediction modeling which make it significantly slower for small programs. However, the dynamic computation cost which corresponds to the number of speculative traces explored by the tool and increases by the size of the program is low.

\noindent $\bullet$ Table~\ref{tab:specStates} shows the number of speculative traces generated for each program on each configuration. It presents a lower-bound on the number of states generated by state-of-the-art tools (No PL), the number of speculative traces generated by Specognitor (with PL), and the number of common speculative traces between them. For PHT-based configurations, Specognitor soundly reduces the number of speculative traces by up to $\sim\!48\%$ for {\ttfamily DES} program (on average by $\sim\!22\%$ for all programs). Moreover, for BTB-based configurations, only a few common speculative traces exist.

\begin{table}
\centering
\caption{\small Execution time of Libgcrypt benchmark.}
{\tiny\sffamily
\setlength\tabcolsep{4pt}
\begin{tabular}{| c | c c | c c | c c | c |}
\hline
\multirow{2}{*}{Libgcrypt} & \multicolumn{2}{c |}{PHT:4} & \multicolumn{2}{c |}{BTB:8} & \multicolumn{2}{c |}{PHT:4,BTB:8} & \multirow{2}{*}{KLEE} \\
 & $\omega=16$ & $\omega=32$ & $\omega=16$ & $\omega=32$ & $\omega=16$ & $\omega=32$ & \\
\hline
hash & $0.83$ & $1.75$ & $0.1$ & $0.11$ & $1.33$ & $2.11$ & $0.02$ \\
des & $0.16$ & $0.39$ & $0.03$ & $0.1$ & $0.12$ & $0.32$ & $0.01$ \\
str2key & $61.9$ & $66.31$ & $65.45$ & $61.77$ & $58.94$ & $63.69$ & $58.3$ \\
camellia & $13.55$ & $14.14$ & $12.62$ & $13.51$ & $16.69$ & $21.75$ & $12.01$ \\
salsa20 & $0.15$ & $0.42$ & $0.04$ & $0.04$ & $0.1$ & $0.23$ & $0.01$ \\
seed & $10.15$ & $8.64$ & $9.59$ & $9.64$ & $9.73$ & $10.42$ & $7.84$ \\
aes & $0.66$ & $1.18$ & $0.42$ & $0.78$ & $0.54$ & $1.05$ & $0.07$ \\
encoder & $4.69$ & $4.65$ & $0.39$ & $0.4$ & $4.69$ & $4.63$ & $0.29$ \\
chacha20 & $1.18$ & $2.36$ & $0.18$ & $0.32$ & $1$ & $2.04$ & $0.03$ \\
ocb3 & $0.14$ & $0.81$ & $0.14$ & $0.18$ & $0.34$ & $0.55$ & $0.1$ \\
\hline
\end{tabular}}
\label{tab:exeTime}
\end{table}

\begin{table}
\centering
\caption{\small Speculative states generated for Libgcrypt benchmark.}
{\tiny\sffamily
\setlength\tabcolsep{3pt}
\begin{tabular}{| c | c | c  c | c  c | c  c |}
\hline
\multirow{2}{*}{Libgcrypt} & \multirow{2}{*}{Config.} & \multicolumn{2}{c |}{PHT:4-16} & \multicolumn{2}{c |}{BTB:8} & \multicolumn{2}{c |}{PHT:4-16,BTB:8} \\
 & & $\omega=16$ & $\omega=32$ & $\omega=16$ & $\omega=32$ & $\omega=16$ & $\omega=32$\\
\hline
\multirow{3}{*}{hash} & No PL & $390$ & $400$ & $445$ & $448$ & $2285$ & $2326$ \\
					  & with PL & $365$ & $375$ & $13$ & $14$ & $1124$ & $1164$ \\
					  & Common & $365$ & $375$ & $0$ & $0$ & $365$ & $375$\\
\hline
\multirow{3}{*}{des} & No PL & $281$ & $488$ & $217$ & $316$ & $308$ & $545$ \\
					  & with PL & $148$ & $282$ & $53$ & $94$ & $164$ & $307$ \\
					  & Common & $148$ & $282$ & $0$ & $0$ & $161$ & $302$ \\
\hline
\multirow{3}{*}{str2key} & No PL & $2650$ & $4107$ & $2380$ & $2512$ & $3374$ & $5583$ \\
					  & with PL & $1675$ & $2687$ & $252$ & $282$ & $1752$ & $2831$ \\
					  & Common & $1675$ & $2687$ & $0$ & $0$ & $1751$ & $2830$ \\
\hline
\multirow{3}{*}{camellia} & No PL & $491$ & $1176$ & $674$ & $1346$ & $4772$ & $11527$ \\
					  & with PL & $460$ & $1019$ & $112$ & $662$ & $2355$ & $6822$ \\
					  & Common & $460$ & $1019$ & $0$ & $0$ & $460$ & $4843$ \\
\hline
\multirow{3}{*}{salsa20} & No PL & $386$ & $924$ & $207$ & $217$ & $526$ & $1288$ \\
					  & with PL & $338$ & $789$ & $22$ & $27$ & $381$ & $868$ \\
					  & Common & $338$ & $789$ & $1$ & $1$ & $338$ & $789$ \\
\hline
\multirow{3}{*}{seed} & No PL & $99$ & $139$ & $645$ & $679$ & $1090$ & $1216$ \\
					  & with PL & $82$ & $116$ & $49$ & $65$ & $285$ & $397$ \\
					  & Common & $82$ & $116$ & $0$ & $0$ & $82$ & $116$ \\
\hline
\multirow{3}{*}{aes} & No PL & $1004$ & $1289$ & $969$ & $1790$ & $1220$ & $1567$ \\
					  & with PL & $851$ & $1061$ & $409$ & $991$ & $852$ & $1062$ \\
					  & Common & $851$ & $1061$ & $0$ & $0$ & $851$ & $1061$ \\
\hline
\multirow{3}{*}{encoder} & No PL & $79$ & $89$ & $106$ & $124$ & $190$ & $296$ \\
					  & with PL & $45$ & $55$ & $39$ & $51$ & $136$ & $146$ \\
					  & Common & $45$ & $55$ & $19$ & $19$ & $45$ & $55$ \\
\hline
\multirow{3}{*}{chacha20} & No PL & $1500$ & $2623$ & $823$ & $995$ & $1679$ & $3025$ \\
					  & with PL & $1121$ & $1663$ & $145$ & $234$ & $1192$ & $1783$ \\
					  & Common & $1121$ & $1663$ & $8$ & $14$ & $1121$ & $1667$ \\
\hline
\multirow{3}{*}{ocb3} & No PL & $1038$ & $2133$ & $983$ & $1071$ & $1186$ & $2350$ \\
					  & with PL & $929$ & $1460$ & $109$ & $144$ & $936$ & $1467$ \\
					  & Common & $929$ & $1460$ & $1$ & $3$ & $929$ & $1460$ \\
\hline
\end{tabular}}
\label{tab:specStates}
\end{table}



\section{Discussion}
\label{sec:discussion}
\noindent \textbf{Soundness.} Specognitor's pattern detection mechanism is sound w.r.t. the terminology of program analysis for PHT-based configurations. In other words, when Specognitor reports no vulnerability for these configurations, there is no vulnerability as described by the pattern description mechanism w.r.t. the two-level prediction logic. Since Specognitor explores all feasible traces and extends the speculative traces, it does not miss any vulnerability. However, it does not mean that all vulnerabilities reported by Specognitor are reproducible on a real system since we over-approximate the system behavior (e.g., in modeling $\omega$) to ensure soundness. Therefore, Specognitor presents an efficient sound over-approximation technique in detecting spectre variant 1 w.r.t. the two-level prediction logic.

\noindent \textbf{Limitations and Future Work.} Specognitor accepts programs in LLVM bitcode format which results in extensive applications for different instruction sets that is one of the main goals of this work. However, since LLVM language is strongly typed, we cannot exhibit the full capabilities of the BTB. For example, LLVM does not allow jumping in the middle of another function because of function frames. Therefore, in such cases, Specognitor does not generate speculative traces. Thus, Specognitor is not sound for BTB-based configurations, however, it is quite precise and can report the execution traces that were affected by jumping to the middle of another function. This work can be extended to cover other micro-architectural models that correspond to spectre variant 3 and 4 such as \textit{Return Stack Buffer} (RSB) and \textit{Store To Load} (STL) dependencies. Another future work is to lift binary information to overcome the limitations of BTB-based configurations.


\section{Related Work}
\label{sec:relatedwork}
Research in the field of cache side-channels has gained pace in the recent years, especially with the advent of Spectre attacks \cite{kocher2020spectre}.
We build Specognitor based on symbolic execution. There are some previous attempts that used symbolic execution to address cache side-channel attacks. CacheD \cite{wang2017cached} is an approach based on symbolic execution which looks at a concrete program trace to find a memory access in which the address depends on a secret. Another symbolic execution-based approach is CaSym \cite{brotzman2019casym} which tries to prove absence of cache side-channel by using abstract cache models. However, neither CacheD nor CaSym can detect speculative cache side channels.
A few recent studies use symbolic execution for speculative cache side-channel detection. Spectector \cite{guarnieri2020spectector} provided a security definition called \textit{speculative noninterference} (SNI). It uses symbolic execution to detect violations of SNI in a program. However, it does not allow any key-dependent branching. In addition, it only analyzes x86 binary programs which limits its applicability to other architectures. \textsc{KLEESpectre} \cite{wang2020kleespectre} extended symbolic execution with cache modeling and speculative execution. It can only detect Bounds Check Bypass (BCB) vulnerabilities, that is a subclass of spectre variant 1 attacks. In addition, it does not consider prediction logic, hence explores infeasible speculative paths and might report false positive vulnerabilities. A recent approach that addressed speculative side-channel detection is SpecSafe \cite{brotzman2021specsafe}. This approach introduces CaSym \cite{brotzman2019casym} to the speculative world. It also improved the security definition proposed by Spectector \cite{guarnieri2020spectector} and proposed \textit{speculative aware noninterference} (SANI). However, it cannot model a precise prediction unit and only detects spectre variant 1 cache side-channels.

\section{Conclusion}
\label{sec:conclusion}
In this work, we explained importance of considering prediction logic in detecting vulnerabilities resulting from speculative behavior of programs containing conditional branches and indirect branches/calls. We also proposed dynamically defined patterns to account for existing and future security threats. We built Specognitor that is a novel prediction-aware symbolic execution tool that can model cache behavior, model speculative execution, model prediction logic, and detect vulnerable patterns with a monitor. We showed the effectiveness of these models in detecting leakage in small programs as well as real-world cryptographic benchmarks.

\section{Acknowledgements}
We gratefully acknowledge Dr. Eric Gerard Rothstein Morris for his support towards our research.
\bibliography{ref}
\bibliographystyle{ieeetr}

\end{document}